\documentclass[preprint,aps,prb,superscriptaddress]{revtex4}
\usepackage{graphicx}
\begin{document}

\title{Variational approach to the scattering of charged particles
by a many-electron system}

\author{V. U. Nazarov}

\affiliation {Department of Physics and
 Institute for Condensed Matter Theory,
 Chonnam National University,
 Gwangju 500-757, Korea}

\author{S. Nishigaki}
\affiliation {Kyushu Institute of Technology, Sensui-cho 1-1,
Tobata, Kitakyushu 804-8550, Japan}

\author{J.~M.~Pitarke}

\affiliation {Materia Kondentsatuaren Fisika Saila, Zientzi
Fakultatea,
Euskal Herriko Unibertsitatea, 644 Posta Kutxatila, E-48080 Bilbo,
Basque
Country, Spain}

\affiliation {Donostia International Physics Center and Centro
Mixto CSIC-UPV/EHU, Donostia, Basque Country, Spain}

\author{C. S. Kim}

\affiliation {Department of Physics and
 Institute for Condensed Matter Theory,
 Chonnam National University,
 Gwangju 500-757, Korea}

\date\today

\begin{abstract}
We report a variational approach to the nonlinearly screened
interaction of charged particles with a many-electron system. This
approach has been developed by introducing a modification of the
Schwinger variational principle of scattering theory, which allows
to obtain nonperturbative scattering cross-sections of moving
projectiles from the knowledge of the linear and quadratic
density-response functions of the target. Our theory is
illustrated with a calculation of the energy loss per unit path length
of slow antiprotons moving in a uniform electron gas, which shows
good agreement with a fully nonlinear self-consistent Hartree
calculation. Since available self-consistent calculations are
restricted to low heavy-projectile velocities, we expect our theory
to have novel applications to a variety of processes where
nonlinear screening plays an important role.
\end{abstract}

\maketitle

The interaction of external charges with solid targets is a problem
of both fundamental and practical interest in contemporary
physics.~\cite{Ibach-82,Echenique-90} Since the targets of interest,
either bulk solids, surfaces, or structures of lower dimension
typically
consist of many interacting electrons, a remarkable progress has been
achieved along the lines of the so-called 'dielectric' perturbative
formulation
of scattering. This formulation utilizes density-response functions of
many-body systems to characterize their dynamical reaction to external
perturbations.~\cite{Echenique-90,Liebsch}

Linear-response theory provides a qualitative description of the
energy loss per unit path length of external particles of charge
$Z_1$ interacting with solid targets, i.e., the so-called stopping
power (SP) of the solid. However, it yields SP that is
proportional to $Z_1^2$ and cannot, therefore, account for the
existing differences between the ranges of positive and negative
pions~\cite{Barkas-56} and the slowing of protons and
antiprotons.~\cite{Moller-02s,Moller-04s} The quadratic-response
treatment
considerably improves the description of the interaction of
external charges with solid targets,~\cite{Sung-83,Hu-88} and yields a
$Z_1^3$
correction to the energy loss that accurately accounts for the
measured energy loss of protons and antiprotons in the
high-velocity regime.~\cite{Pitarke-93,Pitarke-95} Considerable
progress has been achieved recently in the framework of
quadratic-response theory;~\cite{Gaztelurrutia-01,Nazarov-02}
however, at low velocities this theory is only able to
quantitatively account for the strong influence of unit-charge
projectiles ($Z_1=\pm 1$) in the case of high-density targets.

Another approach to investigate the interaction of external charges
with a many-electron system is based on the ordinary formulation of
potential-scattering theory. In this
approach,~\cite{Echenique-81,Echenique-86,Nagy-89,Ladanyi-92,Krakovsky-95}
SP for a
heavy particle is determined in the low-velocity limit from the
knowledge of the scattering phase shifts corresponding to a static
nonlinearly screened potential. These potential-scattering calculations
include all orders
in the projectile charge; however, they have the
limitation of being restricted to low velocities ($v<<v_F$, $v_F$
being the Fermi velocity) of heavy projectiles moving in bulk
materials.

In the case of few-body scattering, variational methods of the
Schwinger type are known to provide nonperturbative
representations of the collision matrix,~\cite{Joachain} which in
the case of the scattering of external charges would not be
restricted to low velocities and heavy projectiles. However, the
direct application of these methods to solid targets composed with a
large number of electrons is not feasible, since it requires the
knowledge of the  many-particle ground and excited states of the
target.

In this paper, we put forward a theory which reconciles the
dielectric formulation of scattering with the variational method.
To do this, we introduce a modification of the Schwinger
variational principle of scattering theory to provide a
nonperturbative representation of the scattering cross-section of
external charges interacting with a many-electron system. The
novelty of this method is that it does not require the knowledge
of the many-particle ground and excited states of the target;
instead, only the linear and quadratic density-response functions
are needed, which have been previously
obtained in the framework of time-dependent
density-functional theory (TDDFT).~\cite{Petersilka-96}

Because of the quite general nature of this approach, it can be
readily applied to the investigation of a variety of processes
involving the inelastic scattering of charged particles by
many-electron targets, such as the SP for moving ions,
electron and positron energy-loss spectroscopy, inelastic
low-energy electron diffraction, and hot-electron dynamics. As our
formula only requires the knowledge of the two
lowest-order density-response functions of the target, the
implementation of this approach has a computational cost equal to
that of available quadratic-response theories.

Let us consider the scattering of a charge by an
arbitrary many-electron target, in which the target is scattered
between its ground $|0\rangle$ and
excited  $|n\rangle$ states.  We assume that the initial and final
states $\Psi_i=|{\bf p}_i,0\rangle$ and $\Psi_f=|{\bf
p}_f,n\rangle$ of the projectile-target interacting system contain
a free particle of charge $Z_1$ and momentum ${\bf p}_i$ and ${\bf
p}_f$, respectively. According to the  bilinear form of the
Schwinger variational principle,~\cite{Joachain} the functional
\begin{eqnarray}
\left[ T_{fi} \right]
\hspace{-0.3 mm}
=
\hspace{-0.3 mm}
\langle\Psi_f^{-}|V|\Psi_i\rangle
\hspace{-0.8 mm}
+
\hspace{-0.8 mm}
\langle\Psi_f|V|\Psi_i^{+}\rangle
\hspace{-0.8 mm}
-
\hspace{-0.8 mm}
\langle\Psi_f^{-}|V
\hspace{-0.8 mm}
-
\hspace{-0.8 mm}
V G_0^{(+)}V|\Psi_i^{+}\rangle
\label{Schwinger}
\end{eqnarray}
is stationary under the variation of the trial (unknown)
eigenstates $\Psi_f^{+}$ and $\Psi_i^{-}$ of the full interacting
hamiltonian and at its stationary point gives the {\it exact}
transition-matrix elements between the (known) initial $\Psi_i$
and final $\Psi_f$ states. Here $V$ is the Coulomb interaction
between the projectile and the target, and $G_0^{(+)}$ is the
Green function associated to the hamiltonian of the
projectile-target system without the mutual interaction $V$.

The fractional form of the Schwinger variational principle is
obtained by substituting (see Ref.~\onlinecite{Joachain}, p. 412)
\begin{eqnarray}
\Psi_i^{+} \rightarrow A \Psi_i^{+}, \Psi_f^{-} \rightarrow B
\Psi_f^{-} \label{trial}
\end{eqnarray}
in Eq.~(\ref{Schwinger}) and treating the coefficients $A$ and $B$
as variational parameters. If one then returns to
Eq.~(\ref{Schwinger}) and uses the free states $\Psi_{i,f}$ for the
trial functions, one finds
\begin{eqnarray}
T_{fi} = T^{(1)}_{fi} / \left(1 -T^{(2)}_{fi}/T^{(1)}_{fi}
\right), \label{SchwingerF}
\end{eqnarray}
where $T^{(1)}_{fi}=\langle\Psi_f|V|\Psi_i\rangle$ and
$T^{(2)}_{fi}=\langle\Psi_f|V G_0^{(+)}V|\Psi_i\rangle$ represent
the first two Born transition amplitudes.

Equation~(\ref{SchwingerF}) has proven very useful in atomic
scattering, where the $|0\rangle$ and $|n\rangle$ states of the
target can be known, at least approximately. However, in the case
of solid targets the
many-body ground and excited states  are difficult to know. In
order to find a representation of the scattering cross-section of
external charges interacting with a many-electron system in terms
of density-response functions, which can be calculated from the
knowledge of one-electron states, we first construct the
differential cross-section for the projectile to be scattered
between the free-particle states of momenta ${\bf p}$ and  ${\bf
p}-{\bf k}$ (we use atomic units throughout)~\cite{Joachain}
\begin{eqnarray}
d \sigma / d {\bf
k}=(16 \pi^4 / v) \, \sum\limits_{n}|T_{fn,i0}|^2 \,
\delta(\omega - \omega_{n0}), \label{sigma}
\end{eqnarray}
where the sum is extended over a complete set of eigenstates of
the target, $\omega_{n0}$ are the excitation energies of the
target, $\omega={\bf k}\cdot{\bf v}-{\bf k}^2/2M$,
$M$ is the projectile mass, and ${\bf v}={\bf p}/M$ is the projectile
initial velocity.
We then introduce the functional
\begin{eqnarray}
&&\left[ d \sigma / d {\bf k }\right]=(16 \pi^4/ v)\,
\sum\limits_{n} |\langle\Psi_{fn}^{-}|V|\Psi_{i0}\rangle +
\langle\Psi_{fn}|V|\Psi_{i0}^{+}\rangle \nonumber \\ &&-
\langle\Psi_{fn}^{-}|V -
V G_0^{(+)}V|\Psi_{i0}^{+}\rangle|^2 \, \delta(\omega - \omega_{n0}),
\label{func}
\end{eqnarray}
we compare this functional with that of Eq.~(\ref{Schwinger}), and
noting that
a linear combination of the squares of the absolute value of a
stationary quantity is a stationary quantity, we conclude that the
functional of Eq.~(\ref{func}) gives the {\it exact} cross-section of
Eq.~(\ref{sigma}) at its stationary point with respect to the
variation of the
trial states $\Psi_{i0}^{+}$ and $\Psi_{fn}^{-}$. The variational
principle based upon the functional of Eq.~(\ref{func}) is the
'global' analog (summed over all final states of the target) of the
corresponding 'local' variational
principle based on the functional of Eq.~(\ref{Schwinger}) for the
transition
to a particular state. These two principles are both exact and
equivalent to
each other.

In order to obtain the fractional form of the global variational
principle, we first apply the substitution (\ref{trial})
to the functional (\ref{func}), and we then approximate the trial
functions by the free states. Hence, the problem reduces to finding the
stationary value of the functional
\begin{eqnarray}
&&\hspace{-0.3 cm} \left[ d \sigma / d {\bf k} \right] = 16 \pi^4
/ v \, \times \nonumber
\\
&&\hspace{-0.3 cm}\sum\limits_{n} |(A+B-A\, B) T^{(1)}_{fn,i} + A
\, B \,T^{(2)}_{fn,i} |^2 \, \delta(\omega - \omega_{n0})
\label{F0}
\end{eqnarray}
with respect to the variational parameters $A$ and $B$.
Equation (\ref{F0}) can be rewritten as
\begin{eqnarray}
&&\left[ d \sigma / {d {\bf
k}}  \right] = (A+B-A B)^2 R_2 + \nonumber \\
&&A B (A+B-A B) R_3 +A^2 B^2 R_4.\label{AB1}
\end{eqnarray}
Here, $R_2$, $R_3$, and $R_4$ represent contributions to the
differential cross-section that are proportional to $Z_1^2$,
$Z_1^3$, and $Z_1^4$, respectively,
which can be obtained from the knowledge of the corresponding
density-response functions of the target, as shown
in Ref.~\onlinecite{Nazarov-02}.

In what follows, we neglect the last term of Eq.~(\ref{AB1}) (see
the discussion below).
Solving the variational problem for the function of Eq.~(\ref{AB1})
with respect to the parameters $A$ and $B$, we find
its stationary value as
\begin{equation}\label{result}
d \sigma / d {\bf k}= R_2\,f\left(R_3 /R_2 \right),
\end{equation}
where
\begin{eqnarray}\label{efe}
&&f(x)=\left[16 - 32 x - 56 x^2 + 72 x^3 - 27 x^4\right.\nonumber\\
&&\left.+(2-x)( 4  - 4 x + 9 x^2)^{3/2}\right]/\,32\,(1-x)^3.
\end{eqnarray}
Equation~(\ref{result}) is the 'global' analog of the 'local'
Eq.~(\ref{SchwingerF}). If the quantity $x=R_3/R_2$ is small, we may
expand
the expression (\ref{efe}) and obtain
\begin{eqnarray}
d \sigma / d {\bf k} = R_2 + R_3+\dots,
\label{Born2}
\end{eqnarray}
which agrees with the Born series through second order in the
interaction between the projectile and the target.

\begin{figure}[h]
\includegraphics[width=0.475\textwidth,height=0.31\textwidth]{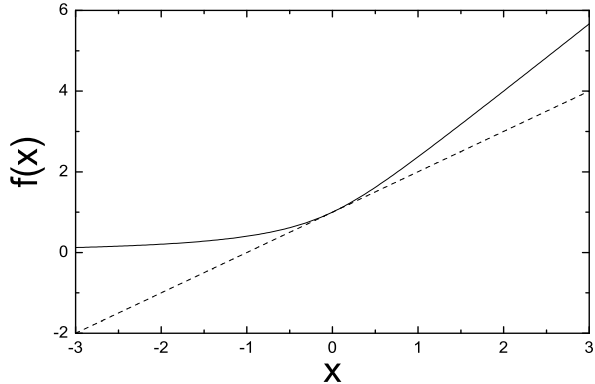}
\caption{\label{Fig1} $f(x)$ of Eq.~(\ref{efe}) (solid line). If $x$
is small, $f(x)=1+x+\dots$ (dashed line), which yields
Eq.~(\ref{Born2}).}
\end{figure}

In the case of antiprotons ($Z_1=-1$) moving slowly in a metal,
the $Z_1^3$ contribution to the scattering cross-section $R_3$,
which is negative, can overcome for low-density metallic targets
the positive $Z_1^2$ contribution $R_2$, thereby
leading to a physically meaningless negative scattering
cross-section. In contrast, our nonperturbative differential
cross-section (\ref{result}) is positively defined, as
shown in Fig.~\ref{Fig1} where the function $f(x)$ of
Eq.~(\ref{efe}) is plotted.

We illustrate our theory with a calculation of the SP
[$-(dE/dx)$] for slow protons and antiprotons
moving in a uniform electron gas (EG) of density $n_0$ characterized
by the
density
parameter $r_s=(3/4\pi)n_0^{-1/3}$. At low velocities the
projectile-target Coulomb interaction is relatively strong, so
this represents an unfavorable situation for a linear or quadratic
perturbative approach.

The SP is obtained by multiplying the differential
cross-section
($d\sigma/d{\bf k}$) by the energy transfer $\omega$ and integrating
over the momentum transfer ${\bf k}$. As the mass $M$ of our
projectile is much larger than the electron mass, $\omega\sim{\bf
k}\cdot{\bf
v}$ and one writes
\begin{eqnarray}
- \ d E / d x = (1/\Omega) \int d{\bf k}\,{\bf v}\cdot{\bf
k}\,(d\sigma/d{\bf k}),
\end{eqnarray}
where $\Omega$ is the normalization volume. We have evaluated the
differential cross-section of Eq.~(\ref{result}) from
the knowledge of the linear and quadratic density-response
functions of the uniform EG, which we have calculated in
the random-phase approximation (RPA).~\cite{Pitarke-95} For slow
projectiles ($v<v_F$), the SP is found to be
proportional to the projectile velocity $v$.

\begin{figure}[h]
\includegraphics[width=0.475\textwidth,height=0.3375\textwidth]{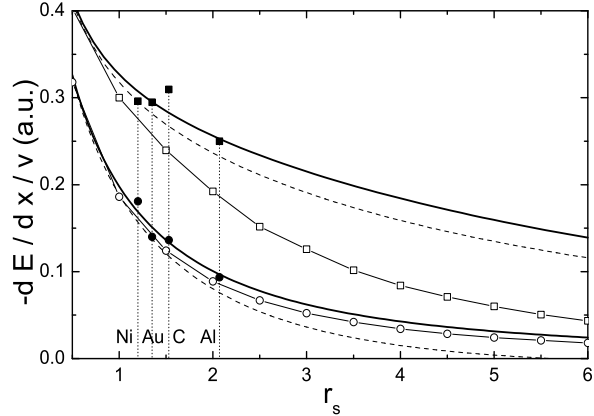}
\caption{\label{Fig2} Variational (solid lines) and quadratic (dashed
lines) SP of a uniform EG for slow antiprotons (lower
curves) and protons (upper curves),
divided by the projectile velocity, as a function of $r_s$. The
chained lines with circles (squares) represent fully nonlinear Hartree
calculations for antiprotons (protons),
performed along the lines of Refs.~\onlinecite{Echenique-86} and
\onlinecite{Nagy-89}
but
with xc excluded. The
circles (squares) show the experimental results for antiprotons
(protons) reported in Refs.~\onlinecite{Moller-02s} and
\onlinecite{Moller-04s}.}
\end{figure}

In Fig.~\ref{Fig2}, the variational SP of a uniform
EG for slow protons and antiprotons (thick solid lines) is compared to
the corresponding quadratic SP [obtained from
Eq.~(\ref{Born2})] (dashed lines) and accurate fully nonlinear
potential-scattering calculations that we have carried out along the
lines of Refs.~\onlinecite{Echenique-86} and \onlinecite{Nagy-89} but with
exchange-correlation (xc)
excluded (chained lines).
In the high-density limit ($r_s\to 0$), the Born series quickly
converges and all calculations coincide. However, as the electron
density decreases, a correct description of nonlinear interactions
requires to go beyond quadratic-response theory.

In the case of antiprotons, quadratic-response calculations (lower
dashed line of Fig.~\ref{Fig2}) overestimate the negative nonlinear
contribution to the SP (see also Fig.~\ref{Fig1}), which
for $r_s>5.5$ becomes larger in magnitude than the linear
($Z_1^2$) term. In contrast, the variational SP (lower
thick solid line) is positive for all electron densities and shows
good agreement with
the fully nonlinear potential-scattering calculation that we have
performed by solving self-consistently the Hartree equation of a
static antiproton (chained line with circles). Since the Hartree
SP is obtained from accurate phase-shift calculations to
all orders in
$Z_1$, the agreement between variational and Hartree calculations
gives us confidence in the accurateness of the variational
approach.

In the case of protons, our variational SP (upper thick
solid line of Fig.~\ref{Fig2}) overestimates the
potential-scattering self-consistent Hartree calculations (chained
line with squares) even more than within quadratic-response theory
(upper dashed line). This is due to the fact that in its
present (initial) form our theory is only applicable to point
charges which do not support bound states or
resonances.\footnote{This difficulty can be overcome by replacing
the plane-wave trial wave functions that we have used in our
variational calculations by wave functions appropriate to the
description of composite projectiles.} Protons in a uniform EG are
known to support either a bound state (at $r_s>2$) or a resonance
below the Fermi level,\cite{Echenique-86} which can be inferred by
the behavior of our calculated scattering phase shifts. Electrons
accommodated in either a bound state or a resonance efficiently
screen the projectile, thereby decreasing the SP as
shown in Fig.~\ref{Fig2}.

The first unambiguous evidence for a velocity-proportional
electronic SP of solid targets has been reported
recently, with measurements of the energy loss of
slow antiprotons in Ni, Au,
C, and Al.~\cite{Moller-02s,Moller-04s} At low
velocities, the
energy loss of ions in metals is mainly due to the slowing by
valence electrons, which can be approximately described with
the use of a uniform EG.~\cite{Campillo-98} Hence, we
have also plotted in Fig.~\ref{Fig2} the friction coefficients
$-(dE/dx)/v$ obtained from the slow-antiproton energy-loss
measurements of Refs.~\onlinecite{Moller-02s} and
\onlinecite{Moller-04s}, at
$r_s=$1.20, 1.35, 1.53, and 2.07 corresponding to the average density
of valence $3d^84s^2$, $5d^{10}6s^1$, $2s^22p^2$, and $3s^23p^1$
electrons in Ni,
Au, C, and Al, respectively. The variational and Hartree calculations
for antiprotons are considerably closer to experiment (solid circles)
than their quadratic counterpart. Our variational calculations for protons are
also
close to experiment; however, this might be due to a cancelation of
the impact of bound and resonance states with the effect of xc
 not included in our approach.
The xc effects were included in
Refs.~\onlinecite{Echenique-81}, \onlinecite{Echenique-86}, and
\onlinecite{Nagy-89}
in the framework of density-functional theory (DFT),\cite{Kohn-65}
and they can be included in our variational procedure by going beyond
the RPA in the description of the density-response functions of the
solid. Work in this direction utilizing the recently obtained quadratic density-response function
from TDDFT with xc included\cite{Nazarov-04} is now in progress.

Although an additional approximation of omitting the last term in
Eq.~(\ref{AB1}) has been made, we note that by including it a more
general analytical formula for the differential cross-section
would be obtained. However, while the terms $R_2$ and $R_3$
require the knowledge of the linear and
quadratic density-response functions, respectively, which are
available from the literature for a number of systems of physical
importance, $R_4$ is related to the cubic response function,
which has not been determined as yet. Although
the generalization of the present theory in this direction is
conceptually straightforward, for the sake of simplicity (and
implementability in calculations at present) we defer it until
later publications.

In conclusion, we have reported a new nonperturbative variational
approach to the nonlinearly screened interaction of charged particles
with a many-electron system, which goes beyond the conventional linear
and quadratic theories. This approach has been developed by
introducing a modification of the Schwinger variational principle
of scattering theory, which allows to obtain nonperturbative
scattering cross-sections from the knowledge of the linear and
quadratic density-response functions of the target. Our approach,
which includes contributions to all orders in the
projectile-target Coulomb interaction and agrees with the Born
series through second order, represents a considerable improvement
over the quadratic theory ($Z_1^3$) approximation; in particular, our
variational
differential cross-section is positively defined, which is known
not to be the case for the quadratic theory.

We have illustrated our theory with a calculation of the stopping power
of uniform EG for slow protons and antiprotons,
which in the case of antiprotons shows good agreement with fully
nonlinear
Hartree calculations. Our calculations indicate that
by going beyond quadratic theory the variational procedure
considerably improves the
agreement with recent measurements of the stopping power of Ni, C,
Al, and Au for slow antiprotons, though xc effects still need to be
taken into account. In the case of protons, the presence of bound
states and resonances also needs to be incorporated into our theory.
Since  self-consistent Hartree
calculations have the limitation of being restricted to low
heavy-projectile velocities ($v<<v_F$), we expect our theory to have
novel applications in the
investigation of a variety of processes involving the inelastic
scattering of charges by many-electron targets.

We thank P. M. Echenique and E. Zaremba for helpful discussions.
V.U.N. and C.S.K. acknowledge support by the Korea Research
Foundation through Grant No. KRF-2003-015-C00214.
J.M.P. acknowledges partial support by the UPV/EHU, the
Basque Hezkuntza, Unibertsitate eta Ikerketa Saila, and the Spanish
MCyT.

\end{document}